\documentclass[12pt,a4paper,twoside]{article}
\usepackage{hyperref}

\def\TheAuthor{Ananda Roy and David P. DiVincenzo}                                   % Author name
\def\TheTitle{Topological Quantum Computing}                        % Title of contribution
\def\TheHeading{Topological Quantum Computing}                                % Short running title
\def\TheInstitute{Institute for Quantum Information, RWTH Aachen University and \newline Peter Gr\"unberg Institut and Institute for Advanced Simulation, Forschungszentrum J\"ulich GmbH}              % Institute
\def\TheAddress{}            % Place e.g. University ...
\def\ThePart{D}                                             % Part of book
\def\TheChapter{7}                                          % Chapter of contribution

\usepackage{ferienschule_15}                                  % Style file for layout

\makeindex
\begin{document}

\MakeTitel           %%% Displays title, author name, etc.
\tableofcontents     %%% Displays table of contents

\vfill
\rule{42mm}{0.5pt}\\
{\footnotesize Lecture Notes of the $48^{{\rm th}}$ IFF Spring
School ``Topological Matter -- Topological Insulators, Skyrmions \newline and Majoranas''
(Forschungszentrum J{\"{u}}lich, 2017). All rights reserved. }

\newpage

%%%%%%%%%%%%%%%%%%%%%%%%%%%%%%%%%% Add text here ... %%%%%%%%%%%%%%%%%%%%%%%%%%%%%%%%%%%%%%%%%%%%%

\section{Introduction}
\label{intro}
Quantum computers have shown great promise as a resource providing exponential speedup over classical computers for certain problems. Akin to their classical counterparts, quantum computers will be prone to errors. There are two major sources of errors. First, the most formidable source of error is decoherence. To perform quantum computation, quantum information \iffindex{quantum information} must be stored, processed and read out while protecting it from the debilitating effects of decoherence. Second, even if the system is protected from decoherence, it is almost certain that all the operations performed on the quantum information during its processing will be imperfect. These errors will accumulate over the duration of the computation, eventually causing failure. Thus, for a quantum computing scheme to be feasible, it needs to be fault-tolerant. What this means is that the quantum computer can still perform its task effectively while its components are imperfect. A major breakthrough in this field is the threshold theorem, which asserts that ideal quantum circuits can be simulated efficiently by noisy ones provided the error rates of individual gates are below a certain threshold \cite{Shor_1996, Aharonov_Ben-Or_1996, Knill_Zurek_1998}. 

Topological quantum computing is an approach to fault-tolerant quantum computation, where the protection from errors occurs not from active intervention, but at a hardware level. The advantage of this approach is that it is robust to localized imperfections. First envisioned by Alexei Kitaev \cite{Kitaev_2003}, \iffindex{Kitaev} this scheme makes use of a non-abelian \iffindex{non-abelian} exchange statistics of elementary excitations of some two-dimensional quantum many-body system. These excitations form the degenerate ground-space of this topologically ordered system, and all other states are separated from this space by a finite excitation gap. In order to perform computation in this scheme, one needs to create pairs of these non-abelian excitations (called non-abelian anyons) \iffindex{anyon} from vacuum, separating them spatially, transporting them around each other to implement the logical gates and finally, fusing them together for measurements. 

In order for this scheme to be viable, there are two conditions that must be fulfilled. First, there could be errors due to quantum tunneling between the non-abelian anyons. These processes can take place even at zero temperature. But, the amplitude of these processes are exponentially suppressed with the spatial separation of these particles and goes as $e^{-L/l_0}$, where $L$ is the spatial separation and $l_0$ is some characteristic length scale of topological ordering of the system. Thus, to prevent such tunneling during quantum information processing and storage, $L$ has to be much larger than $l_0$. Second, at finite temperature, undesired thermally excited quasiparticles can lead to errors. The generation rate of these particles are exponentially suppressed by a Boltzmann factor $e^{-\Delta/T}$, where $\Delta$ is the excitation energy and $T$ is the temperature. Therefore, the temperature must be low enough so that there are sufficiently low number of these undesired excitations. 

There are several physical systems where these excitations could, in principle, be found, manipulated and measured. Non-abelian anyons were proposed as elementary excitations of the $\nu=5/2$ state in the fractional quantum hall effect \cite{Moore_Read_1991}. The braiding statistics of these particles and their possible use in quantum computing has been investigated thoroughly. For more information on this topic, the reader is invited to consult \cite{Stern_2008, Nayak_DasSarma_2008}. In a pathbreaking work \cite{Read_Green_2000}, it was proposed that the non-abelian statistics of the excitations of the $\nu=5/2$ state are shared by the Majorana \iffindex{Majorana} excitations of a 2D spinless $p+ip$ superconductor. Since then, there have been other theoretical models \cite{Levin_Wen_2005, Kitaev_2006} which also support these non-abelian excitations. Moreover, several experimental realizations of these Majorana excitations in solid state systems have also been proposed (for a comprehensive review, see \cite{Alicea_2012}). 

In this lecture note, we will focus on implementation of topological quantum computation with Majorana fermions. The note is organized as follows. We begin by reviewing the basics of abelian anyons in Sec. \ref{basic_abel_anyon}. Then, we describe how abelian anyons can be found in a theoretical model, namely the honeycomb model, \iffindex{honeycomb model} proposed by Alexei Kitaev \cite{Kitaev_2006}. In Sec. \ref{hc_model}, we describe the model, followed by a brief outline of its exact solution and finally describe the abelian anyonic excitations present in this model. Next, we cover the basics of Majorana fermions in Sec. \ref{maj_basic}, followed by a demonstration of their non-abelian exchange statistics in Sec. \ref{nonabel_maj}. Subsequently, in Sec. \ref{maj_example}, we describe how these excitations can occur in the honeycomb model mentioned above in presence of a magnetic field. Next, we describe how to perform topological quantum computing with these Majoranas. We keep our discussion sufficiently general so that they can applied to any system that supports these excitations. In Sec. \ref{cliff_maj}, we describe protocols to implement Clifford gates with these Majorana excitations, followed by implementation of controlled-Z gate in Sec. \ref{cphase_maj} and $\pi/8$ gate in Sec. \ref{pi/8_maj}. We briefly discuss the effect of imperfections in these protocols in Sec. \ref{imperf_maj}. Finally, concluding remarks are presented in Sec. \ref{concl}. 

\section{Abelian anyons}
\label{abel_anyon}
\subsection{Basics of abelian anyons}
\label{basic_abel_anyon}
Abelian anyons are particles which exist in $(2+1)$ dimensions. Exchange of these particles along some topologically specified trajectories gives rise to a multiplication of the overall wavefunction of the quantum state of the system by a phase-factor $e^{i\varphi}$. In general, $\varphi$ can be any rational multiple of $2\pi$.  Since clockwise and counter-clockwise exchanges are not equivalent, the group of these exchanges is the infinite braid group instead of the permutation group \cite{Kauffman_1993}. Abelian anyons transform as one-dimensional representation of this group. 

The simplest example of an abelian anyon is a flux-charge composite particle that occurs in $(2+1)$ dimensional electromagnetism models, first studied by Wilczek \cite{Wilczek_1982_a, Wilczek_1982_b}. In this model, the charges take integer values and the magnetic vortices carry fluxes which are real numbers. Each of these excitations are separately bosonic, but considered together they show nontrivial statistics due to the Aharonov-Bohm effect \cite{Sakurai_1993}. This is understood as follows. When a charge $q$ goes around a flux $\phi$, the system picks up an overall Aharonov-Bohm phase $2\pi q\phi/h$, where $h$ is the Planck's constant. As in the Aharonov-Bohm effect, this phase is topological in nature, {\it i.e.} it is robust to deformation of the trajectory and depends only on the overall winding number of the charge about the flux. Now, what happens when two of these charge-flux composites $(q,\phi)$ are interchanged? When they are interchanged, each of the charges go half-way around the flux of the other composite. Therefore, each contribute an Aharonov-Bohm phase of $\pi q\phi/h$. Thus, the wave-function gets multiplied by $e^{i\varphi}$, where $\varphi = 2\pi q\phi/h$. Note that this phase is the same as rotating one of these composite particles $(q,\phi)$ by $2\pi$, which is what one expects from the usual spin-statistics relation. 

Next, we describe the exchange statistics of molecules composed of anyons. Consider two molecules composed of $n$ anyons. Let each of the anyons have an exchange statistic given by the phase $\varphi$. What is then the exchange statistic of these two molecules? Due to the interchange, each of the $n$ charges of one molecule goes around $n$ fluxes of the other molecule. Thus, each molecule contributes a phase of $n^2\varphi/2$ and the total wave-function is multiplied by a factor $e^{in^2\varphi}$ (see Chap. 9.4 of \cite{Preskill_notes}).

In the next section, we will describe a physical model for interacting spins on a honeycomb lattice, first proposed by Kitaev \cite{Kitaev_2006}, which supports these abelian excitations for a certain choice of parameters\footnote{As we will see later, this model also supports non-abelian anyons for a different choice of parameters.}. We will see that the abelian anyons in this model are exactly the flux-charge composites described above. 

\subsection{A physical realization of abelian anyons}
\label{hc_model}
\subsubsection*{The honeycomb model}
Consider spin-$1/2$ particles (spin being the only relevant degree of freedom) located at the vertices of a honeycomb lattice (cf. Fig. \ref{fig_honeycomb_1}). They interact through nearest neighbor interaction, and the form of the interaction depends on the type of link that connects them. For instance, two spins connected by an x-link interact with an XX interaction. Similar behavior holds for y-links and z-links. Thus, the Hamiltonian is given by: 
\begin{equation}
\label{hc_ham}
H = -J_x \sum_{x-links}\sigma_j^x\sigma_k^x-J_y \sum_{y-links}\sigma_j^y\sigma_k^y-J_z \sum_{z-links}\sigma_j^z\sigma_k^z,
\end{equation}
where $J_x, J_y, J_z$ are the parameters that determine the strength of these interactions. We will choose these parameters to be positive for simplicity. The case for any of them being negative can be analyzed similarly. 
\begin{figure}
 \centering
     \includegraphics[width=0.9\hsize]{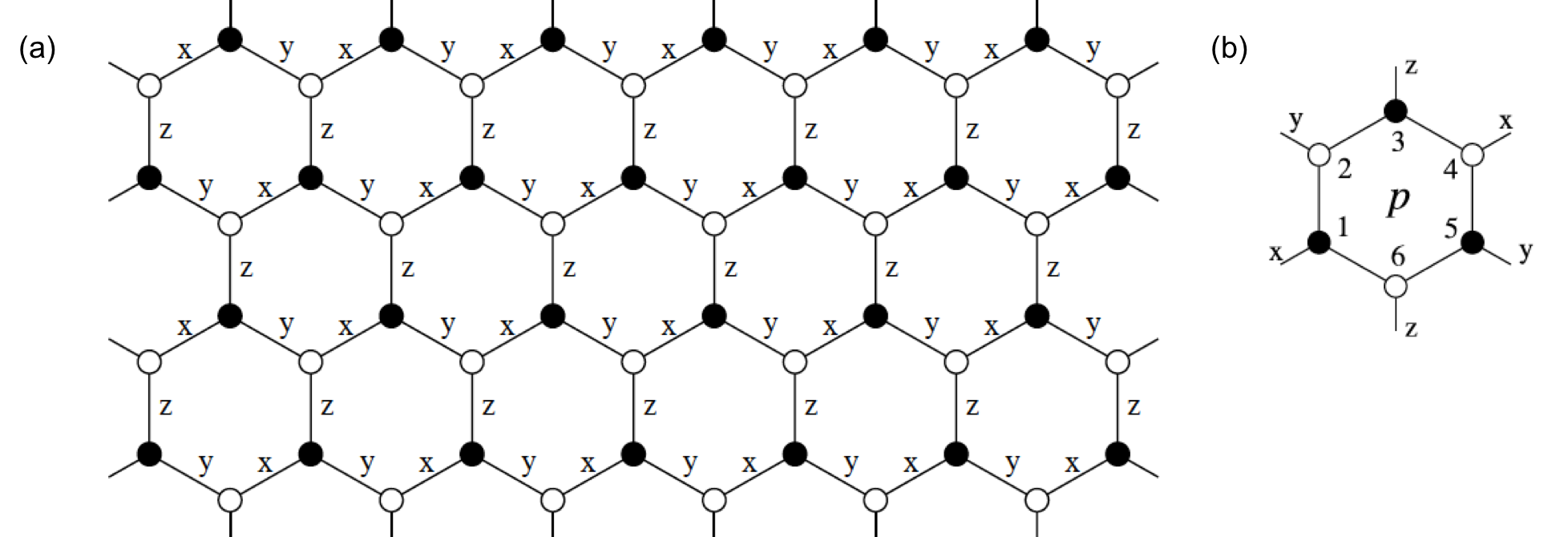}
     \caption{\label{fig_honeycomb_1}(a) Schematic of the honeycomb lattice (image from \cite{Kitaev_2006}). The lattice is composed of two sublattices, shown with solid and empty circles. Spin-$1/2$ particles are located at each vertex of the lattice and they interact through nearest neighbor interaction. The form of the interaction depends on the type of link that connects them. For instance, two spins connected by an x-link interact through an XX interaction. Similar behavior holds for the y-links and z-links. (b) One plaquette of honeycomb lattice (image from \cite{Kitaev_2006}). The plaquette operator ($W_p$) is given by the product of six Pauli operators, one for each site of the plaquette. Which kind of Pauli operator shows up at each site depends on the nature of the external link at each site. }
\end{figure}
One can check that plaquette operators for each hexagonal plaquette defined by $W_p = \sigma_1^x\sigma_2^y\sigma_3^z\sigma_4^x\sigma_5^y\sigma_6^z$ (with eigenvalues $w_p=\pm1$) are conserved quantities since they commute with each other and $H$. Thus, the total Hilbert space splits up into sectors of different plaquette operator eigenspaces. In fact, in each sector, one can exactly solve the problem. This is outlined below. 

\subsubsection*{Exact solution of the honeycomb model}
\label{hc_exact}
There are two distinct ways to solve the honeycomb model. One was proposed by Kitaev in his pioneering work \cite{Kitaev_2006} by mapping each spin to four Majorana fermions. Here, we take an alternate approach, where we solve the problem using Jordan-Wigner transformation. This transformation maps the spins to fermions in an equivalent brick-wall lattice with open boundary condition, with the enumeration of the spins shown in Fig. \ref{fig_honeycomb_2}. Our analysis follows closely that of Chen and Nussinov \cite{Chen_Nussinov_2008}. 
\begin{figure}
 \centering
     \includegraphics[width=0.8\hsize]{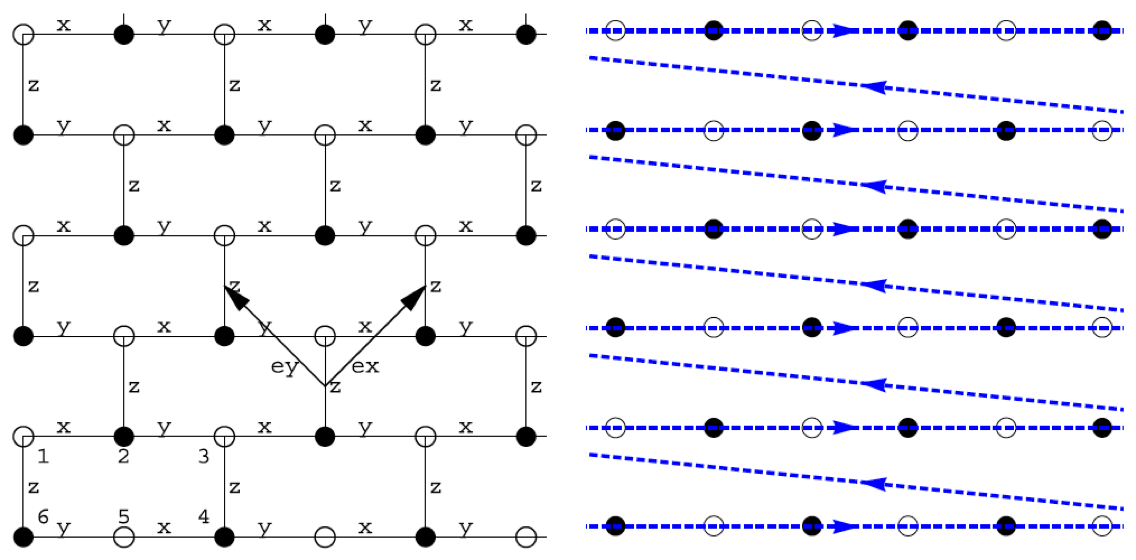}
     \caption{\label{fig_honeycomb_2}Brick-wall lattice, equivalent to the honeycomb lattice (image from \cite{Chen_Nussinov_2008}). In blue is shown the enumeration of the spins used for the Jordan-Wigner transformation.}
\end{figure}

We begin by mapping the spins to fermions as follows: 
\begin{eqnarray}
\sigma_{ij}^+ &=& \Big[\prod_{j'<j}\prod_{i'}\sigma_{i'j'}^z\Big]\Big[\prod_{i'<i}\sigma_{i'j}^z\Big]c_{ij}^\dagger,\\
\sigma_{ij}^z &=& 2c_{ij}^\dagger c_{ij}-1,
\end{eqnarray}
where the indices $i,j$ denote the cartesian coordinates of the lattice sites and $\sigma_{ij}^+ = (\sigma_{ij}^x + i\sigma_{ij}^y)/2$. Under this mapping, the Hamiltonian becomes
\begin{eqnarray}
H &=& J_x\sum_{x-links}(c^\dagger -c)_w(c^\dagger + c)_b-J_y\sum_{y-links}(c^\dagger +c)_b(c^\dagger - c)_w\nonumber\\&&-J_z\sum_{z-links}(2c^\dagger c-1)_b(2c^\dagger c-1)_w,
\end{eqnarray}
where $w,b$ indicate whether the fermion is on a white or black lattice site. We introduce Majorana operators for each of the black and white lattice sites as follows: 
\begin{eqnarray}
A_w = i(c^\dagger -c)_w, B_w = (c^\dagger + c)_w, A_b = (c^\dagger +c)_b, B_b = i(c^\dagger - c)_b.
\end{eqnarray}
Then, the Hamiltonian transforms to: 
\begin{eqnarray}
\label{hc_bcs_1}
H = -iJ_x\sum_{x-links}A_wA_b+iJ_y\sum_{y-links}A_bA_w-iJ_z\sum_{z-links}\alpha_rA_bA_w.
\end{eqnarray}
In the last line, we have defined the operator $\alpha_r \equiv iB_bB_w$ along each z-link, where $r$ is the coordinate of the midpoint of the link. Note that each $\alpha_r$ commutes with the Hamiltonian and is thus a conserved quantity. 

Next, we solve the above Hamiltonian for $\alpha_r = 1, \forall r$. This is the relevant choice to solve the Hamiltonian in the (ground space) vortex-free sector \cite{Kitaev_2006, Chen_Nussinov_2008}, when all the $w_p=1$. To that end, we define a fermion operator as: 
\begin{eqnarray}
d = (A_w + iA_b)/2, d^\dagger = (A_w - iA_b)/2.
\end{eqnarray}
This leads to 
\begin{eqnarray}
H &=& J_x\sum_r (d_r^\dagger+d_r)(d^\dagger_{r+e_x}-d_{r+e_x})+J_y\sum_r (d_r^\dagger+d_r)(d^\dagger_{r+e_y}-d_{r+e_y})\nonumber\\&&+J_z\sum_r(2d_r^\dagger d_r-1),
\end{eqnarray}
where $e_x,e_y$ are unit vectors shown in Fig. \ref{fig_honeycomb_2}. Fourier transforming the above equation yields a p-wave superconducting Hamiltonian: 
\begin{equation}
\label{bcs_p_wave_eqn_1}
H_g = \sum_q \Big\{\epsilon_q d_q^\dagger d_q+\Big(i\frac{\Delta_q}{2}d_q^\dagger d_{-q}^\dagger + h.c.\Big)\Big\},
\end{equation}
where the subscript $g$ indicates that this Hamiltonian describes the vortex free sector. The energy and gap parameters of this superconducting Hamiltonian is given by: 
\begin{equation}
\label{bcs_p_wave_eqn_2}
\epsilon_q = 2J_z-2J_x\cos q_x-2J_y \cos q_y, \ \Delta_q = 2J_x\sin q_x+2J_y \sin q_y.
\end{equation}
The Hamiltonian $H_g$ can easily be diagonalized by Bogoliubov transformation. The detailed form of the excitation spectrum is not relevant for our discussion and the interested reader can consult \cite{Chen_Nussinov_2008} for details. However, one can already see from Eq. \eqref{bcs_p_wave_eqn_2} that the spectrum is gapless for: 
\begin{eqnarray}
\label{eqn_phase_diag}
J_x\leq J_y + J_z, J_y\leq J_z + J_x, J_z\leq J_x + J_y,
\end{eqnarray}
and the system is in a gapped phase whenever these conditions are violated. The complete phase-diagram is shown in Fig. \ref{honeycomb_phase_diag}. 

Next, we will analyze the excitations in one of the gapped phases of the honeycomb model. We will show that these excitations are indeed abelian anyons. To that end, it will be sufficient to consider the model in perturbation theoretical limit $J_x, J_y\ll J_z$ in phase $A_z$. Our analysis will follow that of \cite{Kitaev_2006}.
\begin{figure}
 \centering
     \includegraphics[width=0.6\hsize]{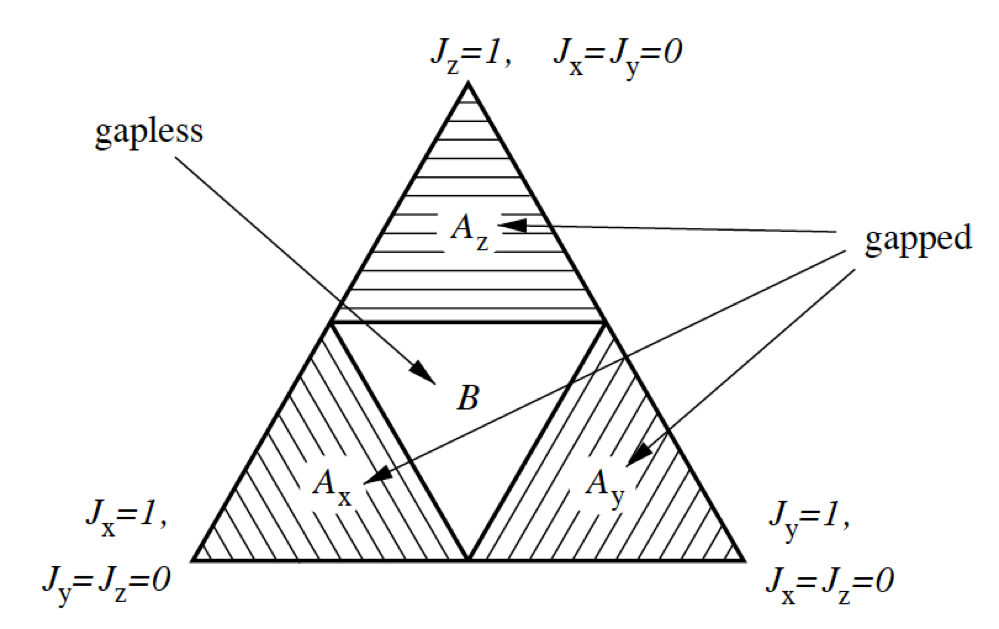}
     \caption{\label{honeycomb_phase_diag}Phase diagram for the honeycomb model (image from \cite{Kitaev_2006}). The model supports three gapped phases ($A_x, A_y, A_z$) and one gapless phase ($B$), depending on the parameters $J_x, J_y, J_z$ [cf. Eq. \eqref{eqn_phase_diag}].}
\end{figure}

\subsubsection*{Abelian anyons in the honeycomb model}
\label{hc_abel}
The starting point of this perturbation theory analysis is the Hamiltonian ($H=H_0 + V$) given in Eq. \eqref{hc_ham}, where 
\begin{equation}
H_0 = -J_z \sum_{z-links}\sigma_j^z\sigma_k^z,\ V = -J_x \sum_{x-links}\sigma_j^x\sigma_k^x-J_y \sum_{y-links}\sigma_j^y\sigma_k^y.
\end{equation}
The ground state for $J_x=J_y=0$ is highly degenerate with each pair of spins connected by z-links being either in $|\uparrow\uparrow\rangle$ or in $|\downarrow\downarrow\rangle$ and can be thought of an effective spin. Next, we include the perturbing Hamiltonian $V$. In order to compute the effective Hamiltonian on this ground space, one needs to perform either a self-energy calculation or a Schrieffer-Wolff transformation (see Appendix B of \cite{Winkler_2003}). The non-trivial Hamiltonian shows up in the fourth-order calculation \cite{Kitaev_2006}. We merely state the final result: 
\begin{equation}
\label{tc_ham}
H_{\rm{eff}} = -J_{\rm{eff}}\Big(\sum_{\rm{vertices}}A_s + \sum_{\rm{plaquettes}}B_p\Big),
\end{equation}
where $J_{\rm{eff}} = J_x^2J_y^2/(16J_z^3)$ and $A_s, B_p$ are defined below. Here, we have already made a unitary transformation that puts each of the effective spins of the original model on the links of a square lattice (Fig. \ref{toric_code_fig}). Note that this is in contrast to the original model where the spins are on the vertices. The operators $A_s, B_p$ are defined as: 
\begin{equation}
A_s = \prod_{\rm{star(s)}}\sigma_j^x, \ B_p = \prod_{\rm{boundary(p)}}\sigma_j^z,
\end{equation}
where the $\sigma_j^{x,z}$ denote the Pauli operators for the effective spins. 
\begin{figure}
 \centering
     \includegraphics[width=0.4\hsize]{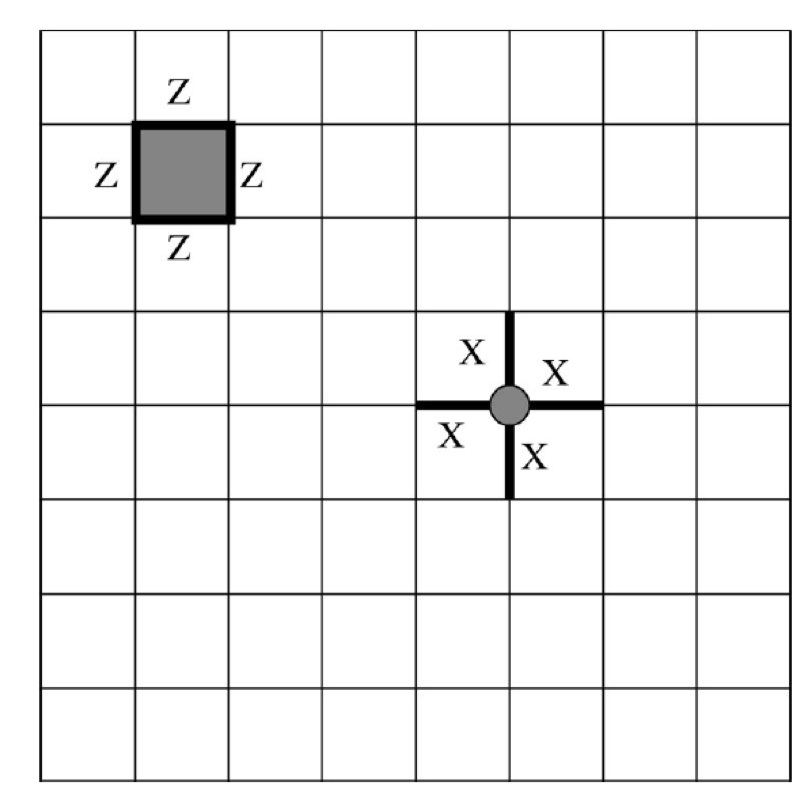}
     \caption{\label{toric_code_fig} Effective lattice for the Hamiltonian given in Eq. \eqref{tc_ham} (image from \cite{Dennis_Preskill_2002}). Here, the effective spins of the honeycomb lattice reside on the links and not on the vertices. The Hamiltonian is given in Eq. \eqref{tc_ham}, which is composed of star operators $A_s = \prod_{\rm{star(s)}}\sigma_j^x$ and the plaquette operators $B_p = \prod_{\rm{boundary(p)}}\sigma_j^z$. }
\end{figure}
It is easy to check that the operators $A_s, B_p$ commute with each other and thus, with $H_{\rm{eff}}$. The ground state is given by the state with the eigenvalues of each of $A_s, B_p$ being $+1$. Excited states can be obtained by flipping the eigenvalues of the star and plaquette operators. When a specific $A_s$ operator has eigenvalue $-1$, the excitation is called an electric charge and is located at the vertex $s$. On the other hand, when a specific $B_p$ operator has eigenvalue $-1$, the excitation is a magnetic vortex and is located at the plaquette $p$. These electric charges and magnetic vortices behave exactly like the ones described in Sec. \ref{basic_abel_anyon}. Both the electric charges and the magnetic vortices are bosonic when considered separately. However, moving an electric charge around a magnetic vortex gives a nontrivial phase of $-1$, as would be expected due to the Aharonov-Bohm effect. Thus, the electric charge and magnetic vortex excitations of the gapped phase are abelian anyons. A more detailed proof of this can be found in \cite{Chen_Nussinov_2008}. The different superselection sectors of this phase and their braiding rules of them are explicitly given in \cite{Kitaev_2006}. 

\section{Non-abelian anyons}
\label{non-abel_anyons}
Non-abelian anyons are excitations in $(2+1)$ dimensions, which when exchanged along some topologically specified trajectories, the overall wavefunction of the system gets multiplied by a unitary matrix. Since matrix-multiplication is non-commutative, these excitations show non-abelian exchange statistics. In terms of the braid group representations, the non-abelian anyons transform according to representations which have dimensions $>1$. In this note, we focus on one kind of such non-abelian particles: the Majorana fermions. 

\subsection{Basic definition of a Majorana fermion}
\label{maj_basic}
In this section, we describe the basic properties of Majorana fermions, following the treatments of \cite{Ivanov_2001, Leijnse_Flensberg_2012}. Consider $2n$ spatially well-separated Majoranas $\gamma_1, \ldots \gamma_{2n}$. Since a Majorana degree of freedom is half a fermionic degree of freedom, one can combine them to give rise to full fermions: 
\begin{eqnarray}\label{eq:fermion}
	f_j &=& (\gamma_{2j-1} + i \gamma_{2j})/2.
\end{eqnarray}
This, in turn, implies: 
\begin{eqnarray}
\label{eq:inversefermion1}
	\gamma_{2j-1} &=& f_j^\dagger + f_j, \\
\label{eq:inversefermion2}
	\gamma_{2j} &=& i(f_j^\dagger - f_j).
\end{eqnarray} 
Note that the Majorana operators are hermitian, $\gamma_j = \gamma_j^\dagger$, satisfying the following anti-commutation relations:  
\begin{eqnarray} \label{eq:anticommutation}
	\{\gamma_j, \gamma_k\} = 2 \delta_{jk}.
\end{eqnarray}
The last equation follows from the usual fermion anti-commutation relations for the operators $f_j$. 
Eq.~(\ref{eq:anticommutation}) implies that $\gamma_j^2 = 1$. It is important to realize that it does not make sense to talk about the occupancy of a Majorana mode. The naively constructed ``Majorana number operator", $\gamma_j^\dagger \gamma_j$ is identically $=1$. Similar set of reasoning proves that $\gamma_j \gamma_j^\dagger=1$. Thus, in the traditional sense, the Majorana mode is empty and filled at the same time. However, it is possible to speak of the number states $|n_j\rangle$, which are eigenstates of the number operator $n_j=f_j^\dagger f_j$, $j = 1,\ldots, n$. In terms of the Majorana fermions, these number operators are given by: 
\begin{eqnarray}
n_j = f_j^\dagger f_j = \frac{1}{2}(1+i\gamma_{2j-1}\gamma_{2j}), j = 1,\ldots,n.
\end{eqnarray}
In general for spatially separated Majorana fermions, the way to re-write them in terms of traditional fermions is non-unique \footnote{However, when two Majorana fermion wave-functions overlap, it is natural to combine them into a traditional fermion.}. Note that the ground space of these $2n$ Majorana fermions is $2^n$-fold degenerate, corresponding to each $n_j$ being equal to zero or one.

\subsection{Non-abelian statistics of Majorana fermions}
\label{nonabel_maj}
In this section, we describe the non-abelian exchange statistics of the Majorana fermions. We keep our treatment sufficiently general so that it is applicable to any system that supports these excitations. 

An essential component of non-abelian statistics is a degenerate ground space, which as discussed above, is true for a system supporting spatially separated Majorana fermions\footnote{In general, the degeneracy of the ground space is lifted for finite separation of the Majoranas. The energy splitting is exponentially suppressed with the spatial separation. We will always assume this splitting to be small enough to be negligible.}. Further, this ground space must be separated from all the excited states by an excitation gap, so that the exchange statistics is well-defined. Then, exchange operations, performed adiabatically compared to the excitation gap, can bring the system from one ground state to another. 

Consider again $2n$ spatially localized Majorana fermions: $\gamma_i, i = 1, \ldots, 2n$. Fixing the initial position of the Majoranas, consider a permutation of the Majoranas. The exchange statistics of these Majoranas is given by a unitary representation of the braid group. This group, denoted by $B_{2n}$, is generated by exchange operations $B_{i,i+1}, i= 1,\ldots,2n-1$ of the neighboring Majoranas labeled by $i$ and $i+1$. These operators satisfy the following relations: 
\begin{eqnarray}
\label{comm_braid}
B_{i,i+1}B_{j,j+1} &=& B_{j,j+1}B_{i,i+1},\hspace{1cm} |i-j|>1,\\
B_{i,i+1}B_{j,j+1}B_{i,i+1}&=&B_{j,j+1}B_{i,i+1}B_{j,j+1}, \hspace{1cm}|i-j|=1.
\end{eqnarray}

Next, we give a simple argument to motivate the explicit representation of the braid operations \cite{Hassler_2014}.  Consider a clockwise exchange of the two Majorana fermions, $\gamma_i$ and $\gamma_{i+1}$ (cf. Fig. \ref{braid}). This is accomplished by acting these operators by conjugation with the unitary operator $B_{i,i+1}$. Let us denote the Majorana operators after the exchange by $\gamma'_i$ and $\gamma'_{i+1}$. Therefore,
\begin{equation}
\gamma'_i = B_{i,i+1}\gamma_iB^\dagger_{i,i+1}, \ \gamma'_{i+1} = B_{i,i+1}\gamma_{i+1}B^\dagger_{i,i+1}.
\end{equation}
Since the position of the two Majoranas are interchanged by this operation, 
\begin{eqnarray}
\gamma'_i = \alpha_i \gamma_{i+1}, \ \gamma'_{i+1} = \alpha_{i+1}\gamma'_{i},
\end{eqnarray}
where $\alpha_i, \alpha_{i+1}\in \Re$ since the Majorana operators are real. Since this local exchange operation does not change the fermion number parity, 
\begin{equation}
-i\gamma_i\gamma_{i+1} = -i\gamma'_i\gamma'_{i+1}.
\end{equation}
This implies that \begin{equation}
\alpha_i\alpha_{i+1} = -1.
\end{equation}
Thus, one of the Majorana fermions picks up a negative sign and the other doesn't. There is a gauge degree of freedom in choosing which of the Majoranas picks up the negative sign. We will work with the convention that 
\begin{eqnarray}
\alpha_i= 1, \alpha_{i+1}=-1. 
\end{eqnarray}
Thus, the result of this exchange operation is
\begin{eqnarray}
\label{eq:braid1}
	\gamma_i &\rightarrow&  \gamma_{i+1},\\ 
\label{eq:braid2}
	\gamma_{i+1} &\rightarrow& - \gamma_i,\\
	\gamma_j&\rightarrow& \gamma_j, \ j\notin\{i,i+1\}.
\end{eqnarray}
and the relevant unitary representation of this braid group transformation is 
\begin{eqnarray}\label{eq:exchange}
	B_{i,i+1} = {\rm{exp}}\Big(-\frac{\pi}{4}\gamma_i\gamma_{i+1}\Big)
=\frac{1}{\sqrt{2}} \left( 1 - \gamma_i \gamma_{i+1} \right).
\end{eqnarray}
Similarly, a anti-clockwise exchange instead results in $\gamma_i \rightarrow  -\gamma_{i+1}$, $\gamma_{i+1} \rightarrow \gamma_i$, which is described by the operator 
$B^{-1}_{i,i+1} ={\rm{exp}}\big(\frac{\pi}{4}\gamma_i\gamma_{i+1}\big)= ( 1 + \gamma_i \gamma_{i+1} ) / \sqrt{2}$. 

\begin{figure}
 \centering
     \includegraphics[width=0.9\hsize]{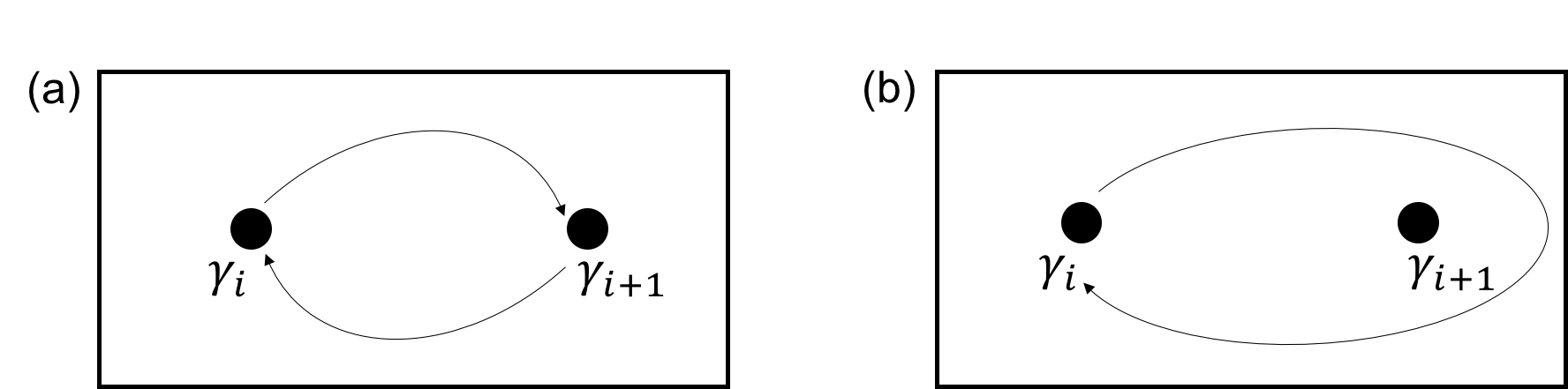}
     \caption{\label{braid}Schematic of braiding operations of Majorana excitations. (a) Schematic of a clockwise exchange of two Majorana fermions $\gamma_i$ and $\gamma_{i+1}$. (b) Schematic of moving the Majorana fermion $\gamma_i$ around $\gamma_{i+1}$ in a clockwise direction. }
\end{figure}

Next, we discuss the effect of bringing one Majorana fermion around another and back to its original position (cf. Fig. \ref{braid}). Topologically, 
this is equivalent to two successive exchanges. Thus, the associated operator is given by $B_{i,i+1}^2 = -\gamma_i \gamma_{i+1}$, leading to the transformation
\begin{eqnarray}
\label{eq:exchangesquare1}
	\gamma_i \rightarrow \left( -\gamma_i \gamma_{i+1} \right) \gamma_i \left( -\gamma_i \gamma_{i+1} \right)^\dagger = -\gamma_i, \\
\label{eq:exchangesquare2}
	\gamma_{i+1} \rightarrow \left( -\gamma_i \gamma_{i+1} \right) \gamma_{i+1} \left( -\gamma_i \gamma_{i+1} \right)^\dagger = -\gamma_{i+1}. 
\end{eqnarray}
Thus, the operation of bringing one Majorana fermion around another results in introducing a minus sign into each Majorana operator. It is easy to check that the operation generated by $B_{i,i+1}^3$ is equivalent to that of $B_{i,i+1}^{-1}$, while $B_{i,i+1}^4$ gives rise to the identity operation. 

Having obtained the explicit representation of the braid group, we are finally in a position to demonstrate the non-abelian statistics of the Majorana fermions. Exchanges of distinct pairs of Majoranas commute [see Eq. \eqref{comm_braid}]. However, whenever two exchanges involve some of the same Majorana fermions, the braid operators do not commute
\begin{eqnarray}\label{eq:noncommuting}
	\left[ B_{i-1,i}, B_{i, i+1} \right] &=& \gamma_{i-1} \gamma_{i+1}.
\end{eqnarray}
The above equation explicitly shows the non-abelian statistics of the Majorana fermions.

\subsection{A physical realization of Majorana fermions}
\label{maj_example}
In this section, we describe a theoretical model whose excitations are these Majorana fermions. To that end, consider again the honeycomb model of Kitaev \cite{Kitaev_2006} described in Sec. \ref{hc_model}. As we saw in the discussion of the exact solution (see Sec. \ref{hc_exact}), the system can be either in a gapped or gapless phase depending on the choice of the coupling constants $J_x, J_y, J_z$ [see Eq. \eqref{eqn_phase_diag}]. Consider the gapless phase [phase $(B)$ of Fig. \ref{honeycomb_phase_diag}] and let us choose $J_x=J_y=J_z=J$ for simplicity. In Sec. \ref{hc_exact}, we showed that this phase supports gapless fermions [cf. Eqs. \eqref{bcs_p_wave_eqn_1},\eqref{bcs_p_wave_eqn_2}] in the vortex-free sector. Vortices can be included by flipping the signs of the variables $\alpha_r$ from $+1$ to $-1$ [see Eq. \eqref{hc_bcs_1}] and they are gapped in all the phases $A_x, A_y, A_z$ and $B$ (see also discussion in Sec. \ref{hc_abel}). However, in phase $B$, because of the gapless fermions, these vortices do not have well-defined exchange statistics {\it i.e.} the transformation of the quantum state depends on the exact trajectory of the exchange. Thus, a spectral gap needs to be opened before one can talk about exchange statistics.  

It can be shown that for this honeycomb lattice, no time-reversal preserving perturbation can open a gap in the $B$ phase (see \cite{Kitaev_2006} for proof). However, adding a uniform magnetic field (which does not preserve time-reversal symmetry) opens the desired gap. We showed in Sec. \ref{hc_exact} that the unperturbed honeycomb model can be mapped to a 2D spinless p-wave superconductor. Here, we will show that adding a magnetic field adds a $ip$ component to the superconductor, in addition to opening a mass gap for the fermions of Sec. \ref{hc_exact}. As a consequence, each vortex then has an unpaired Majorana fermion pinned to it. These Majoranas show non-abelian exchange statistics and can be used for topological quantum computation.

Denoting the different components of the magnetic field as $h_x, h_y, h_z$, we get the total Hamiltonian of the system to be 
\begin{eqnarray}
H_{\rm{tot}} &=& H + H_{\rm{mag}}\nonumber\\
H &=& -J_x \sum_{x-links}\sigma_j^x\sigma_k^x-J_y \sum_{y-links}\sigma_j^y\sigma_k^y-J_z \sum_{z-links}\sigma_j^z\sigma_k^z,\nonumber\\
\label{hc_mag_eq}
H_{\rm{mag}}& =& -\sum_j \big(h_x \sigma_j^x+h_y \sigma_j^y+h_z \sigma_j^z\big).
\end{eqnarray}
Once again, we calculate the effective Hamiltonian on the vortex-free sector. A third order Schrieffer-Wolff or self-energy calculation yields the effective Hamiltonian to be \cite{Kitaev_2006}: 
\begin{equation}
\label{mag_eff_eq}
H_{\rm{mag, eff}} \sim -\frac{h_xh_yh_z}{J^2}\sum_{j,k,l}\sigma_j^x\sigma_k^y\sigma_l^z,
\end{equation}
where the relevant contributing terms are shown in Fig. \ref{hc_mag}(a) and its symmetric permutations. Note that exact prefactor is difficult to compute and this crude estimate of the effective Hamiltonian will be sufficient for our purposes.
\begin{figure}
 \centering
     \includegraphics[width=0.9\hsize]{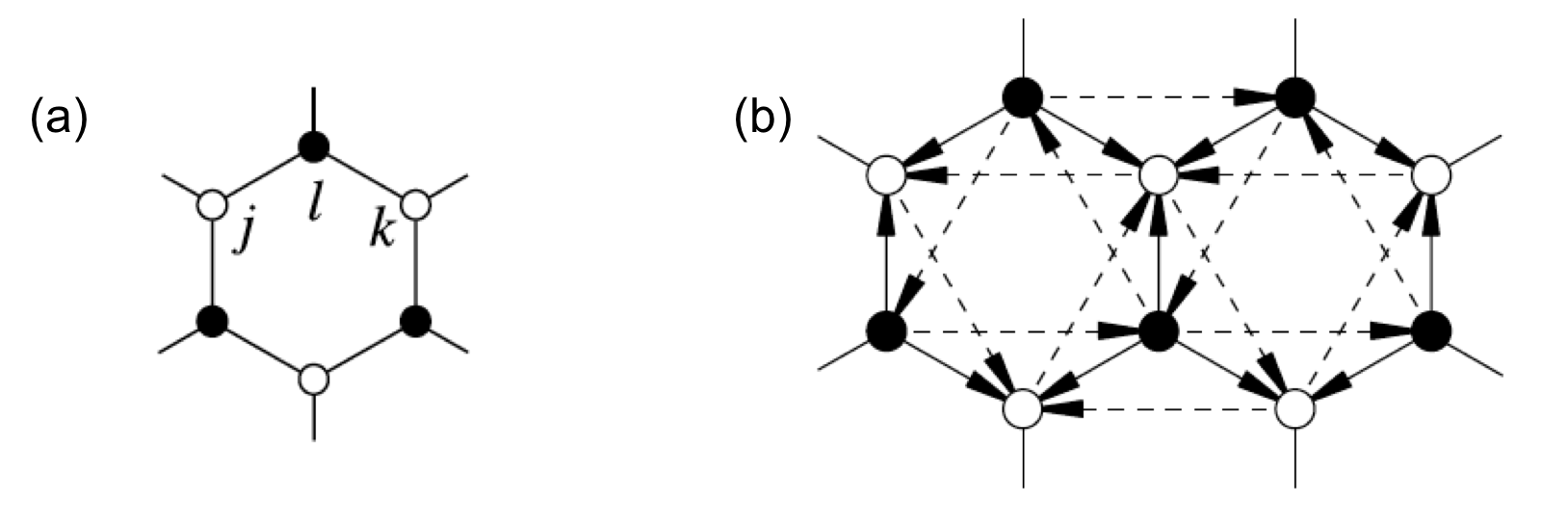}
     \caption{\label{hc_mag} Images from \cite{Kitaev_2006}. (a) Schematic of a contributing term in the effective Hamiltonian due to the magnetic field [see Eq. \eqref{mag_eff_eq}]. (b) Schematic of the interactions of the spins in the honeycomb lattice in presence of a magnetic field. In addition to the nearest neighbor interaction, shown as solid arrows, present due to the unperturbed Hamiltonian [Eq. \eqref{hc_ham}], the magnetic field gives rise to next-to-nearest neighbor interaction [see Eq. \eqref{hc_mag_eq}], shown as dashed arrows. This next-to-nearest neighbor interaction is the crucial ingredient that opens the gap in the spectrum.}
\end{figure}
In the perturbation calculation, the total Hamiltonian is then given by 
\begin{eqnarray}
\tilde{H}_g &=& -J_x \sum_{x-links}\sigma_j^x\sigma_k^x-J_y \sum_{y-links}\sigma_j^y\sigma_k^y-J_z \sum_{z-links}\sigma_j^z\sigma_k^z-\frac{h_xh_yh_z}{J^2}\sum_{j,k,l}\sigma_j^x\sigma_k^y\sigma_l^z.
\end{eqnarray}
In addition to the nearest neighbor interaction between the spins present in the unperturbed Hamiltonian, the magnetic field adds next-to-nearest neighbor interactions [see Fig. \ref{hc_mag}(b)]. To analyze this Hamiltonian, we will again use the Jordan-Wigner transformation method \cite{Chen_Nussinov_2008} outlined in Sec. \ref{hc_exact}. One could also use the original method of Kitaev \cite{Kitaev_2006} by mapping each spin to four Majorana fermions. Since most of the steps are similar to what is described in Sec. \ref{hc_exact}, we merely state the final effective Fourier-transformed Hamiltonian 
\begin{equation}
\tilde{H}_g = \sum_q \Big\{\epsilon_q d_q^\dagger d_q+\Big(i\frac{\Delta_q + i\tilde\Delta_q}{2}d_q^\dagger d_{-q}^\dagger + h.c.\Big)\Big\},
\end{equation}
where 
\begin{eqnarray}
\label{bcs_p+ip_wave_eqn_2}
\tilde\Delta_q &=& \frac{4h_xh_yh_z}{J^2}\big\{\sin(q_y-q_x) + \sin q_x -\sin q_y\big\}
\end{eqnarray}
and $\epsilon_q, \Delta_q$ are defined earlier. In the vicinity of the Dirac points of the unperturbed Hamiltonian $(q_x, q_y) = (\pm \pi/3, \mp \pi/3)$, one can expand the above Hamiltonian and show that $\tilde{H}_g$ indeed supports the same Majorana excitations as a 2D spinless $p+ip$ superconducting Hamiltonian. The Majorana edge modes appear at the boundary between regions with $h_xh_yh_z>0$ and  $h_xh_yh_z<0$ (more on Majorana wavefunctions for a $p+ip$ superconductor can be found in \cite{Alicea_2012}). Moreover, vortex excitations in this phase have Majorana zero modes attached to them \cite{Burnell_Nayak_2011}. The proof in terms of the Chern number and more details on the superselection sectors and braiding rules can be found in \cite{Kitaev_2006}. 

Next we describe how to perform universal quantum computation using these Majorana fermions. We will keep our discussion sufficiently abstract and general so that they can be applied to any physical system that supports these excitations.

\section{Quantum computing with Majorana fermions}
\subsection{Clifford operations on Majorana qubits}
\label{cliff_maj}
In this section, we describe a computational model based on Majorana fermions. Following \cite{Bravyi_2006}, first we define the computational Hilbert space, in which one can prepare an initial state, a set of unitary operations on this Hilbert space and a set of measurements. 

In principle, it is possible to encode a qubit in two Majorana fermions. The ground and excited states of the qubit will then be the unoccupied and occupied states of the fermionic mode of Eq. \eqref{eq:fermion}. However, due to fermion superselection rules, one cannot prepare this qubit in a superposition of ground and excited states. Therefore, we will redundantly encode a qubit in 4 Majorana fermions. 
Thus, the computational Hilbert space of $n$ qubits, $\mathbb{C}^{2n}$, will be encoded in $4n$ Majorana fermions $\gamma_i$, $i = 1, \ldots, 4n$.  We choose the logical subspace to be given by the constraint $\gamma_{4i-3}\gamma_{4i-2}\gamma_{4i-1}\gamma_{4i}=-1$, $i=1,\ldots,n$. In this space, the initial state: $|\bm{0}\rangle = |0\rangle^{\otimes n}$ can be generated by preparing quadruples of these $4n$ Majorana fermions from vacuum. The connection between the logical Pauli operators of the $n$ qubits and the $4n$ Majorana fermions can be written as: 

\begin{eqnarray}
\sigma^{(i)}_z &=& -i\gamma_{4i-3}\gamma_{4i-2}\\\sigma^{(i)}_x &=& -i\gamma_{4i-2}\gamma_{4i-1},\\\sigma^{(i)}_y &=& -i\gamma_{4i-3}\gamma_{4i-1}.
\end{eqnarray}
As shown in Sec. \ref{nonabel_maj} [cf. Eq. \eqref{eq:exchange}], the nearest neighbor exchange operations are  
\begin{eqnarray}
B_{i,i+1} = {\rm{exp}}\Big(-\frac{\pi}{4}\gamma_i\gamma_{i+1}\Big).
\end{eqnarray}
These nearest neighbor exchanges can be composed to give rise to a nonlocal exchange operation
\begin{equation}
 B_{i,j}={\rm{exp}}\Big(-\frac{\pi}{4}\gamma_i\gamma_{j}\Big), i\leq j-2
\end{equation} as follows:
\begin{equation}
B_{i,j} = B_{j-1,j}\cdots B_{i+1,i+2}B_{i,i+1}B_{i+1,i+2}^\dagger\cdots B_{j-1,j}^\dagger.
\end{equation}
Here, the nonlocal exchange operator $B_{i,j}$ acts on the Majorana fermions in the following manner
\begin{eqnarray}
B_{i,j}\gamma_k B_{i,j}^\dagger &=& \gamma_k,\ {\rm{if}}\ k\notin\{i,j\},\nonumber\\&=&\gamma_j,\ {\rm{if}}\ k=i,\nonumber\\&=&-\gamma_i,\ {\rm{if}}\ k=j.
\end{eqnarray}
Among the set of measurements, the nearest neighbor fusion process of two Majorana fermions gives rise to a non-destructive projective measurement of the observable \begin{equation}
F_{i,i+1} = -i \gamma_i\gamma_{i+1}.
\end{equation}
These, together with the braiding operations, give rise to measurements of any observable $F_{i,j}$ because
\begin{equation}
F_{i,j} = B_{i+1,j}F_{i,i+1}B_{i+1,j}^\dagger.
\end{equation}

However, performing the aforementioned braiding operations and measurements is not sufficient for performing universal quantum computing. This can be understood as follows. First, using the mapping of the Pauli matrices to Majorana fermions and the conjugating action of the braid operations, it follows that any braid operation maps the group of Pauli operators to itself. Therefore, all the braid operators belong to the Clifford group. Second, the measurement operators involve only Pauli operators. Thus, by Gottesman-Knill theorem \cite{Nielsen_Chuang_2000}, any computation performed by the braid operations and measurements described above can be simulated efficiently by a classical computer.  

An alternate way to understand the same is in terms of fermionic linear optical quantum computing (FLOQC) \cite{Knill_2001, Terhal_DiVincenzo_2002, Bravyi_2005}. In terms of FLOQC, the initial state is the Fock vacuum, the braid operations are canonical transformations generated by quadratic Hamiltonians and the observables $F_{i,j}$ are single-mode occupation numbers. Therefore, these operations can be efficiently simulated classically \cite{Bravyi_2005}. A similar connection can also be made with linear optical quantum computing with Fock states \cite{KLM_2001}.

As will be shown below, this limitation is overcome by a resource of two ancilla states, denoted by $|a_4\rangle$ and $|a_8\rangle$. The first state, $|a_4\rangle$, enables an operation beyond the Clifford group, namely a rotation by $\pi/8$. The second state, $|a_8\rangle$, enables a nonlinear operation beyond FLOQC, namely, the controlled-phase gate, that can be used to entangle qubits. In what follows, we describe how this is accomplished assuming availability of ideal ancilla states. The case of imperfect ancillas will be alluded to at the end. 

\subsection{Implementation of a controlled-phase gate}
\label{cphase_maj}
In this section, we prove, following \cite{Bravyi_2006}, that ancilla qubits (composed of eight Majorana modes) in the state $|a_8\rangle$, together with single-qubit Clifford operations, can be used to perform a controlled-phase rotation on a system qubit (composed of four Majorana modes), where 
\begin{equation}
|a_8\rangle \equiv\frac{1}{\sqrt{2}}\big(|0,0\rangle + |1,1\rangle\big).
\end{equation}
In the first step of the proof, we show that ancilla qubits in the state $|a_8\rangle$, together with braiding operations, can be used to make a nondestructive four-Majorana-mode measurement. In the second step, we show that this four-Majorana-mode measurement allows one to make a four-Majorana-mode unitary operation. Note that this four-Majorana unitary gate cannot be accomplished by simple braiding operations. In the third step, we show that this four-Majorana mode unitary gate, together with single-qubit Clifford operations, give rise to the controlled-phase gate.

Consider a qubit, encoded in four Majorana modes $\gamma_i,i=1,\ldots,4$, in an arbitrary state $|\psi\rangle$ and two ancilla qubits, encoded in eight Majorana modes $\gamma_i, i = 5, \ldots,12$, in the state $|a_8\rangle$. First, the circuit in Fig. \ref{a_8_circ}(a) is applied to the joint state $|\psi\rangle\otimes|a_8\rangle$. Second, the observables $T_1 = -i\gamma_1\gamma_2, T_2= -i\gamma_3\gamma_4, T_3 = -i\gamma_5\gamma_6$ and $T_4 = -i\gamma_7\gamma_8$ are measured. Third, the Clifford gates $\gamma_2\gamma_9, \gamma_4\gamma_{10}, \gamma_6\gamma_{11}$ and $\gamma_8\gamma_{12}$ are applied [not shown in Fig. \ref{a_8_circ}(a)]. It can be shown (see \cite{Bravyi_2006} for details) that these set of actions amount to projecting $|\psi\rangle$ to $1/2(I\pm \gamma_1\gamma_2\gamma_3\gamma_4)|\psi\rangle$, followed by teleportation of the qubit state encoded in $\gamma_i, i = 1,\ldots,4$ to the four Majorana modes $\gamma_i, i = 9, \ldots, 12$ [Fig. \ref{a_8_circ}(b)]. Here, the $\pm$ depends on the measurement outcomes in the second step. This, shows that an ancilla state $|a_8\rangle$, together with single-qubit Clifford operations, gives rise to a four-Majorana projective measurement. This concludes the first step of the proof. 

\begin{figure}
 \centering
     \includegraphics[width=0.9\hsize]{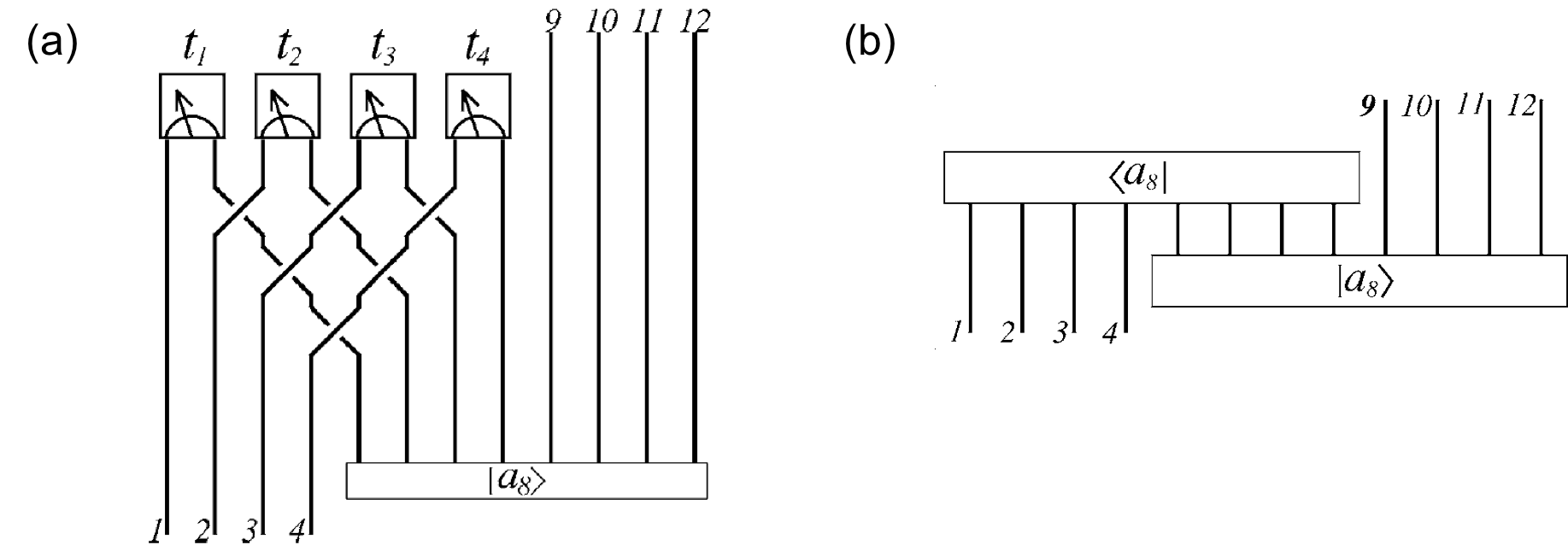}
     \caption{\label{a_8_circ} Images from \cite{Bravyi_2006}. (a) Schematic of the circuit needed for implementation of four-Majorana projective measurement. (b) Teleportation circuit, equivalent to the circuit in (a). }
\end{figure}

For the second step, consider six-Majorana modes $\gamma_i, i = 1, \ldots, 6$ prepared in a state $|\psi'\rangle$, such that $(\gamma_5 + i\gamma_6)|\psi'\rangle=0$. Next, we make a measurement of $\gamma_1\gamma_2\gamma_4\gamma_5$, followed by that of $-i\gamma_3\gamma_5$. It can be shown that these two measurements, together with single-qubit Clifford gates, gives rise to a unitary rotation of ${\rm{exp}}(i\pi/4\ \gamma_1\gamma_2\gamma_3\gamma_4)$ (see \cite{Bravyi_2006} for details). This concludes the second step of the proof. 

For the third step, we need to only show that the four-Majorana mode unitary operation, together with single-qubit Clifford gates, is sufficient to perform a controlled-phase gate. To that end, consider two qubits encoded in the Majoranas $\gamma_i, i = 1,\ldots,4$ and $\gamma_j, j = 5, \ldots, 8$ respectively. Then, the controlled-phase gate is given by: 
\begin{eqnarray}
\Lambda(\sigma_z) = {\rm{exp}}\Big\{i\frac{\pi}{4}(1-\sigma^{(1)}_z)(1-\sigma^{(2)}_z)\Big\},
\end{eqnarray}
where $\sigma_z^{(1)} = -i\gamma_3\gamma_4$ and $\sigma_z^{(2)} = -i\gamma_5\gamma_6$. Thus, 
\begin{eqnarray}
\Lambda(\sigma_z) = e^{i\pi/4}{\rm{exp}}\Big(-i\frac{\pi}{4}\gamma_3\gamma_4\gamma_5\gamma_6\Big){\rm{exp}}\Big(-\frac{\pi}{4}\gamma_3\gamma_4\Big)\Big(-\frac{\pi}{4}\gamma_5\gamma_6\Big).
\end{eqnarray}
Therefore, the controlled-phase gate is indeed composed of a four-qubit unitary rotation ${\rm{exp}}\big(-i\pi/4\ \gamma_3\gamma_4\gamma_5\gamma_6\big)$, together with single-qubit Clifford operations. This completes the proof. 

\subsection{Implementation of a $\pi/8$ rotation}
\label{pi/8_maj}
In this section, we describe the proof that an ancilla qubit in the state $|a_4\rangle$ can be used to perform a rotation by $\pi/8$ on a target qubit in an unknown state following \cite{Bravyi_Kitaev_2005, Bravyi_2006}, where 
\begin{equation}
|a_4\rangle \equiv \frac{1}{\sqrt{2}}\big(|0\rangle + e^{i\pi/4}|1\rangle\big).
\end{equation}
Consider a system qubit in an unknown state $\psi = a|0\rangle + b|1\rangle$. Thus, the system and ancilla qubits together are in the state $|\psi\rangle\otimes|a_4\rangle$. First, perform the joint measurement $\sigma_z\otimes\sigma_z$ on the two qubits. This can be done since any two-qubit Pauli measurement can be reduced to a single-qubit Pauli measurement using Clifford operations. The outcomes for this measurement $\pm1$ appear with probability $1/2$, with the final two-qubit state being projected to
\begin{eqnarray}
|\Psi_1^+\rangle &=& a|0,0\rangle + be^{i\pi/4}|1,1\rangle, \\|\Psi_1^-\rangle &=& ae^{i\pi/4}|0,1\rangle + b|1,0\rangle.
\end{eqnarray}
Second, apply the controlled-not gate on the two qubits with the system qubit as the control qubit. This can be achieved by a combination of the controlled-phase gate (whose implementation was described above) and the single-qubit Clifford rotations. After this step, the state of the two-qubits is given by: 
\begin{eqnarray}
|\Psi_2^+\rangle &=& \big(a|0\rangle + be^{i\pi/4}|1\rangle\big)\otimes|0\rangle, \\|\Psi_2^-\rangle &=& \big(ae^{i\pi/4}|0\rangle + b|1\rangle\big)\otimes|1\rangle.
\end{eqnarray}
In the final step, the ancilla is measured in the $\{|0\rangle, |1\rangle\}$ basis. An outcome of $|0\rangle$ results in a $\pi/8$ rotation on the system qubit, while for an outcome $|1\rangle$, an extra braid gate $|0\rangle\langle0| + i|1\rangle\langle1|$ accomplishes the same. 

\subsection{Robustness to imperfect preparations of ancilla qubits}
\label{imperf_maj}
As explained in Sec. \ref{cliff_maj}, the topologically protected braiding operations and measurements are insufficient to perform universal quantum computing. To accomplish the latter, one needs to have additional ancilla qubits in certain `magic' states ($|a_4\rangle$ and $|a_8\rangle$). Therefore, these magic states are necessarily generated by non-topological operations which are noisy and unprotected. One way to implement these non-topological operations is to bring two anyons sufficiently close to each other, wait for a desired amount of time and then returning the anyons to their initial positions. Since this operation is noisy, instead of having perfect ancilla qubits in states $|a_4\rangle$ and $|a_8\rangle$, one prepares them to some precision, characterized by their fidelity $\epsilon_i = 1-\langle a_i|\rho|a_i\rangle, i = 4,8$, where $\rho$ is the density matrix of the ancilla qubit. A major breakthrough in this field was accomplished with the result of \cite{Bravyi_2006}, which proves that for $\epsilon_4< 0.14$ and $\epsilon_8<0.38$ and perfect topologically protected Clifford operations, one can distill ideal states $|a_4\rangle$ and $|a_8\rangle$. The details of the proof lie outside the scope of this set of lecture notes, but the interested reader is invited to consult \cite{Bravyi_Kitaev_2005, Bravyi_2006} for details. 

\section{Conclusion}
\label{concl}
To summarize, we have presented, in this lecture note, a review of topological quantum computation with Majorana fermions. First, we discussed basic properties of abelian anyons and described a theoretical model that supports these excitations. Second, we discussed the basic properties of Majorana fermions. We discussed their non-abelian exchange statistics and described a theoretical model where these excitations can be found. Third, we discuss how to perform topological quantum computing with the Majorana fermions. We discuss the implementation of single qubit Clifford gates, controlled-Z gate and $\pi/8$ gate, which together are sufficient for universal quantum computing. 

\section{Acknowledgments}
Discussions with Fabian Hassler, Barbara Terhal and Daniel Zeuch are gratefully acknowledged. A.R. acknowledges the support through the ERC Consolidator Grant No. 682726 and D.P.D. acknowledges the support of the Alexander von Humboldt foundation. 

%%%%%%%%%%%%%%% Appendices %%%%%%%%%%%%%%%%%

\newpage

%%%%%%%%%%%%%%% References %%%%%%%%%%%%%%%%%

\end{document}